\newif\ifanon

\ifanon
\documentclass[manuscript,screen,review,anonymous]{acmart}
\else
\documentclass[sigconf,screen]{acmart}
\fi

\usepackage{booktabs}
\usepackage{subcaption}
\usepackage{enumitem}

\copyrightyear{2025}
\acmYear{2025}
\setcopyright{rightsretained}
\acmConference[PETRA '25]{The PErvasive Technologies Related to Assistive Environments}{June 25--27, 2025}{Corfu Island, Greece}
\acmBooktitle{The PErvasive Technologies Related to Assistive Environments (PETRA '25), June 25--27, 2025, Corfu Island, Greece}
\acmPrice{}
\acmDOI{10.1145/3733155.3734896}
\acmISBN{979-8-4007-1402-3/25/06}

\begin{document}
\title{Using Confidence Scores to Improve Eyes-free Detection of Speech Recognition Errors}

\author{Sadia Nowrin}
\email{snowrin@mtu.edu}
\orcid{0000-0003-1321-3868}

\author{Keith Vertanen}
\email{vertanen@mtu.edu}
\orcid{0000-0002-7814-2450}
\affiliation{%
  \institution{Michigan Technological University}
  \city{Houghton}
  \state{Michigan}
  \country{USA}
}

\begin{abstract}
Conversational systems rely heavily on speech recognition to interpret and respond to user commands and queries. Despite progress on speech recognition accuracy, errors may still sometimes occur and can significantly affect the end-user utility of such systems. While visual feedback can help detect errors, it may not always be practical, especially for people who are blind or low-vision. In this study, we investigate ways to improve error detection by manipulating the audio output of the transcribed text based on the recognizer's confidence level in its result. Our findings show that selectively slowing down the audio when the recognizer exhibited uncertainty led to a 12\% relative increase in participants' ability to detect errors compared to uniformly slowing the audio. It also reduced the time it took participants to listen to the recognition result and decide if there was an error by 11\%.  
\end{abstract}


\maketitle

\section{Introduction}
In recent years, there have been notable advancements in Automatic Speech Recognition (ASR) technology, enabling eyes-free interaction \cite{vtyurina_bridging_2019, huang_duiva_2022, ghosh_commanding_2020, fan_just_2021} and improving accessibility for devices without a visual display (e.g.~Amazon Echo, Google Home). Speech recognition can help make interfaces accessible for individuals with motor impairments \cite{pradhan_accessibility_2018, metatla_voice_2019, bilmes_vocal_2005, nowrin_exploring_2022, wagner_programming_2012} as well as those who are blind \cite{azenkot_exploring_2013}. While deep learning models have advanced ASR accuracy \cite{baevski_wav2vec_2020, liu_unsupervised_2022}, real-world ASR performance is often negatively impacted by background noise, speaker variations, and speaker disfluencies \cite{goldwater_which_2010}. Despite efforts to improve recognition accuracy in noisy environments using large language models \cite{wang_wav2vec-switch_2022, weninger_speech_2015}, only a modest relative improvement of 5.7\% was obtained \cite{wang_wav2vec-switch_2022}. Azenkot and Lee \cite{azenkot_exploring_2013} observed that blind users spent a significant amount of time correcting errors when performing speech dictation tasks. Speech error correction involves a two-step process: 1) detecting errors, and 2) correcting errors \cite{vertanen_automatic_2009}. To fully realize the potential of speech recognition technology, accurate error detection and correction are crucial. In this paper, we focus on the first step, error detection.

\subsection{Related work}
Numerous prior studies have investigated using visual feedback to represent a speech recognizer's confidence in its result \cite{vertanen_chi2008, berke_captions, fuji_correction_2016, price_correct_2005, htut_subtitling, shiver_evaluating, piquard_qualitative}. However, visual feedback may not be possible for individuals with visual impairments, for sighted users in situations in which they cannot visually attend to their device, or when using a device without a screen such as a smart speaker. Identifying errors in conversational systems without visual feedback can be challenging for a variety of reasons. Firstly, text-to-speech (TTS) audio can be hard to understand, especially when errors involve short or similar-sounding words \cite{burke_correction_2006}. In a study with sighted users \cite{hong_identifying_2018}, participants missed approximately 50\% of recognition errors when the TTS audio was played at a rate of 200 words per minute (wpm). Understanding TTS can be even more difficult in noisy environments \cite{cohn_perception}. Finally, errors may occur infrequently, lulling users into trusting the recognizer \cite{noyes_errors}. 

Beyond sighted users, blind individuals also face challenges with detecting recognition errors through audio-only feedback. Hong et al.~\cite{hong_review_2020} found no significant difference in ASR error identification between blind users (42\%) and sighted users (38\%). This was despite blind users' extensive experience with synthesized speech. This suggests that experience with synthesized speech alone may not be sufficient for improving error detection, highlighting the need to explore alternative approaches to enhance audio-based feedback.

In this study, we examine how users can detect speech recognition errors through audio-only feedback. Similar to Hong and Findlater \cite{hong_identifying_2018}, we investigate the impact of various TTS manipulations on users' ability to detect ASR errors. Hong and Findlater found error detection improved when TTS was delivered at 200\,wpm, or even slower at 100\,wpm, compared to a higher speech rate of 300\,wpm. In comparison to past work, we investigate adjusting the audio feedback using the speech recognizer's \textit{confidence score}. The confidence score indicates how certain the ASR system is about the accuracy of its result \cite{gillick_probabilistic_1997}. Additionally, we investigate the error detection in both \textit{common phrases} where all words were in-vocabulary, and \textit{challenging phrases} where at least one word was out-of-vocabulary (e.g.~acronyms, proper names).

We studied the effect of confidence scores on eyes-free error detection by testing four audio annotations: default speech rate, slow speech rate, slow speech rate for low confidence recognitions, and playing a beep for low confidence recognitions. We found slowing down the TTS audio based on the confidence score led to 85\% accurate error detection, outperforming uniform slowing of the audio which had a detection accuracy of 76\%. Despite the increased audio length resulting from the selective slowing according to the confidence score, the participants only experienced a slight 7\% increase in the time it took them to review the recognition results compared to the default speech rate condition.

\section{User Study}
The goal of the user study was to investigate whether modulating the audio presentation of the speech recognition results based on a recognizer's confidence score could improve the ability of users to detect errors. 

\subsection{Participants}
We recruited 48 participants (15 female, 32 male) aged from 21--68 (mean=36, sd=11.5) via Amazon's Mechanical Turk, an online crowdsourcing platform. We opted for an online study in order to to ensure participant safety during the COVID-19 pandemic. Participants were compensated at a rate of \$10 (USD) per hour. Participants completed the experiment in 27 minutes on average. All participants self-reported being native English Speakers. 67\% of the participants agreed that they frequently used speech interfaces with 22\% agreeing that computers had difficulty understanding their speech. 

At the end of the study, participants rated various statements using a 7-point scale with one denoting strongly disagree and seven denoting strongly agree. Statements included how easy it was to identify recognition errors in each of the four audio annotation conditions and whether they could anticipate which sentences would likely have errors. See Appendix \ref{sec_questionnaire} for our exact questionnaires. 

\subsection{Study Design}
We employed a within-subject experimental design with four counterbalanced conditions:
\begin{itemize}[leftmargin=5mm]
    \itemsep0em 
    \item \textbf{\textsc{AllNormal} --- } The recognition result was synthesized into speech and played at 200\,wpm. This is similar to the default speaking rate of commercial TTS systems. 
    \item \textbf{\textsc{AllSlow} --- } The result was played at 70\% of the default speaking rate, equivalent to 140\,wpm. This was somewhat faster than the 100\,wpm used by \citet{hong_identifying_2018}. We selected 140\,wpm as a compromise between slowing the speech to help users spot errors and avoiding excessive listening time. 
    \item \textbf{\textsc{UncertainSlow} --- } If the confidence score was below a threshold, the result was played at 140\,wpm.
    \item \textbf{\textsc{UncertainBeep} --- } If the confidence score was below a threshold, a one-second beep tone was played at the beginning followed by the result played at the default speaking rate of 200\,wpm.
\end{itemize}

In the \textsc{UncertainSlow} and \textsc{UncertainBeep} conditions, we used a \textit{confidence threshold} to determine whether to slow the TTS audio or add a beep. To establish this threshold, we conducted a pilot study with 12 participants. The pilot was conducted similar to our main study but used an initial guess for the threshold. We recognized the 480 utterances collected during the pilot  using Google's speech-to-text service.\footnote{\url{https://cloud.google.com/speech-to-text}} We tested different thresholds measuring: 1) the true positive rate (TPR), the proportion of utterances containing one more or recognition errors that were correctly identified as having an error, and 2) the false positive rate (FPR), the proportion of utterances with no errors that were incorrectly identified as having an error. We evaluated the trade-off between the TPR and the FPR at different thresholds with a receiver operating characteristic (ROC) curve. Based on this analysis, we selected a threshold of 0.93, aiming to balance sensitivity (0.85) and specificity (0.75) to detect a high percentage of errors while avoiding too many false positives. 

\subsection{Procedure}
Using a web application, participants first signed a consent form and completed a demographic questionnaire. Participants then read a set of instructions and completed two practice tasks to familiarize themselves with the task. The audio was played at the default speaking rate in the practice trial. At the start of each condition, we provided participants with a description of how the audio annotation worked for the current condition. 

Participants recorded a sentence for each task which was transcribed by Google's speech-to-text service and then synthesized via Google's TTS service.\footnote{\url{https://cloud.google.com/text-to-speech}} Speech Synthesis Markup Language (SSML) was used in the TTS request to slow the speech rate or add a beep. Following a delay caused by the speech-to-text and TTS processing (averaging around four seconds), participants listened to the audio of the recognition result which was generated using a female voice (en-US). Participants were allowed to play the audio only once. 

\begin{figure*}[tb]
\centering
\fbox{\includegraphics[width=13.25cm]{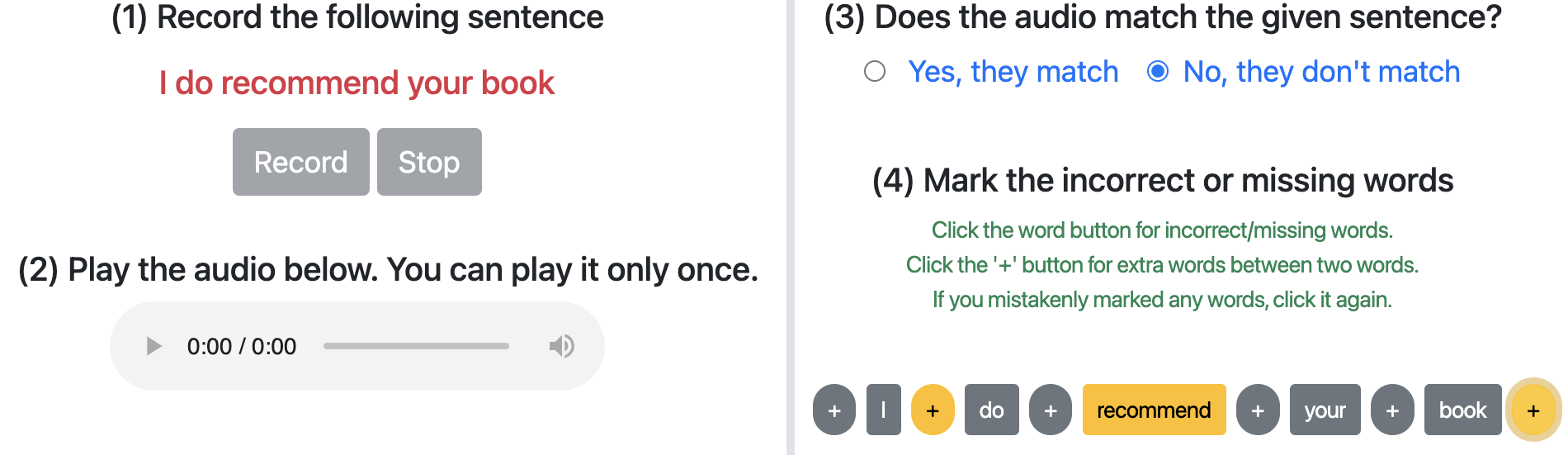}}
\caption{Screenshot of our web application. In part (1), the user records a reference sentence, in this case ``I do recommend your book''.  In part (2), after recognition, a control appears allowing playback of the recognition result. In this case the recognition was ``I really do command your book group''. In part (3), the user specifies if there were any recognition errors. If they answer ``No'', part (4) shows buttons for each word in the reference sentence as well as plus buttons between all words. The word buttons allow the user to specify a word was recognized incorrectly or was missing in the result. The plus buttons allow the user to specify if extra words were recognized in between reference words. Buttons toggled on are highlighted in yellow.}
\label{fig:app}
\Description{A screenshot showing the four phases of the interface used by participants. In phase one, participants record themselves speaking a provided sentence. In phase two, participant could replay the recognized words via text-to-speech. In phase three, they decided if the text-to-speech was correct. In phase four, they marked incorrect or missing words in the recognized words.}
\end{figure*}

After listening to the result, we asked participants if the reference sentence matched the audio. If they answered no, indicating a speech recognition error, we asked them to locate the incorrect or missing words, as well as any incorrect additional words that may have appeared between two words in the reference sentence (Figure \ref{fig:app}). Participants could only detect errors after the audio finished playing, simulating a real-world scenario where users cannot interrupt the system to correct errors during the initial playback. Finally, participants completed a final questionnaire about their experience in each condition and the study as a whole.

We selected phrases from a collection of 407 Twitter phrases \cite{vertanen_velociwatch_2019}. This set included 194 common phrases containing all in-vocabulary words and 213 challenging phrases containing at least one out-of-vocabulary word. Out-of-vocabulary words were those not appearing in a list of 100,000 frequent English words. Challenging phrases included proper nouns and abbreviations that might be difficult for the speech recognizer. We used phrases with 5--10 words. Participants were randomly assigned 40 phrases. Each condition included five common and five challenging phrases that were presented in random order. 

\section{Results}
In total, we collected 1,920 utterances. Google's recognizer had a word error rate (WER) of 15\% on these utterances. Our analysis includes two key measures: the accuracy of the error detection and the detection time. We conducted a one-way repeated measures ANOVA to compare the four conditions. In cases where the normality assumptions were violated (Shapiro-Wilk test, $p$ < 0.05), we employed the non-parametric aligned rank test (ART). We used the Wilcoxon signed-rank test to compare the WER between the common phrases and the challenging phrases.  

\begin{figure}[tb]
    \begin{subfigure}[]{0.35\textwidth}
    \centering
    \includegraphics[width=\textwidth]{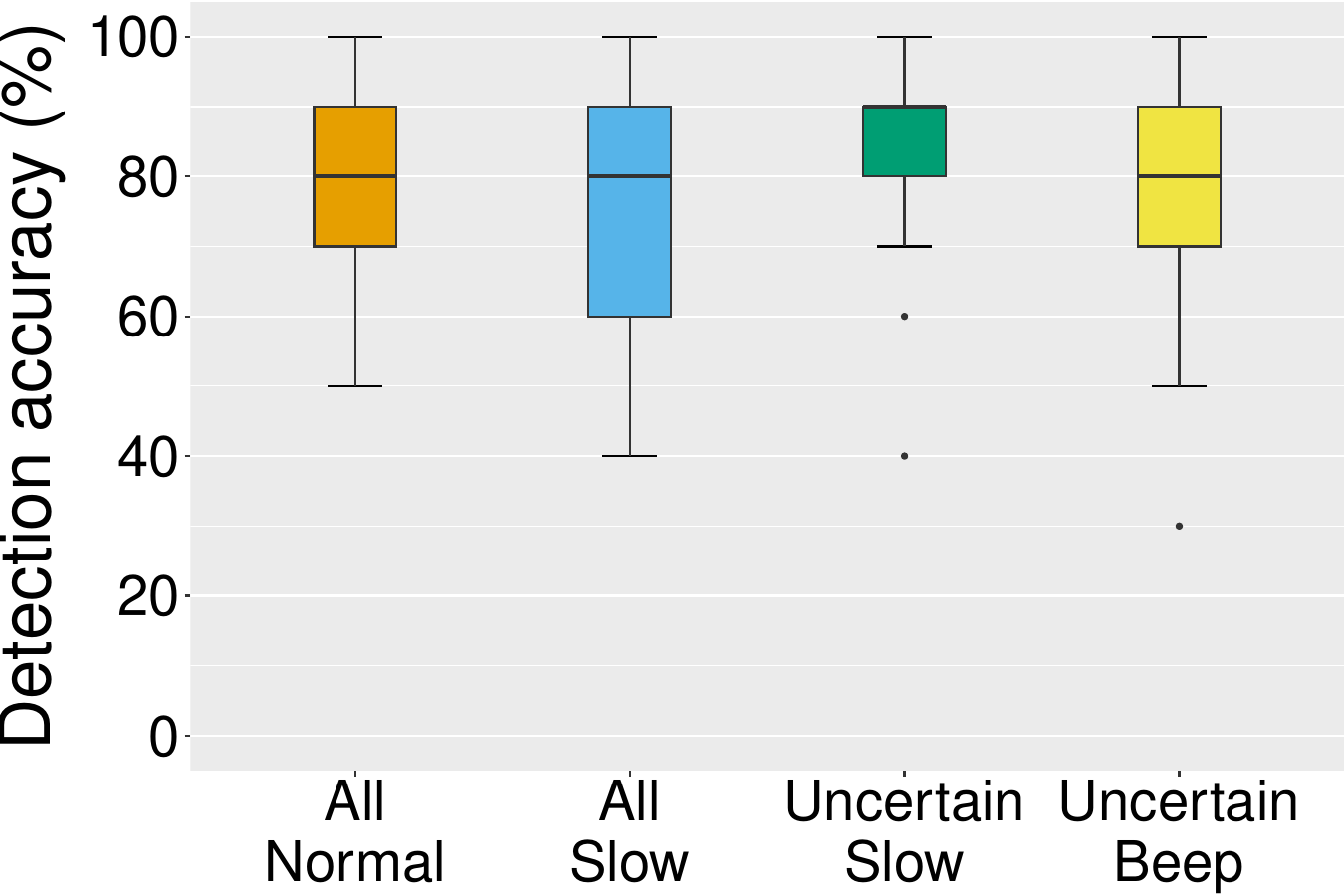} 
    \caption{Accuracy detecting the presence of errors}
    \label{fig:acc}
    \end{subfigure}%
        \vspace{3mm}
    \begin{subfigure}[]{0.35\textwidth}
    \centering
    \includegraphics[width=\textwidth]{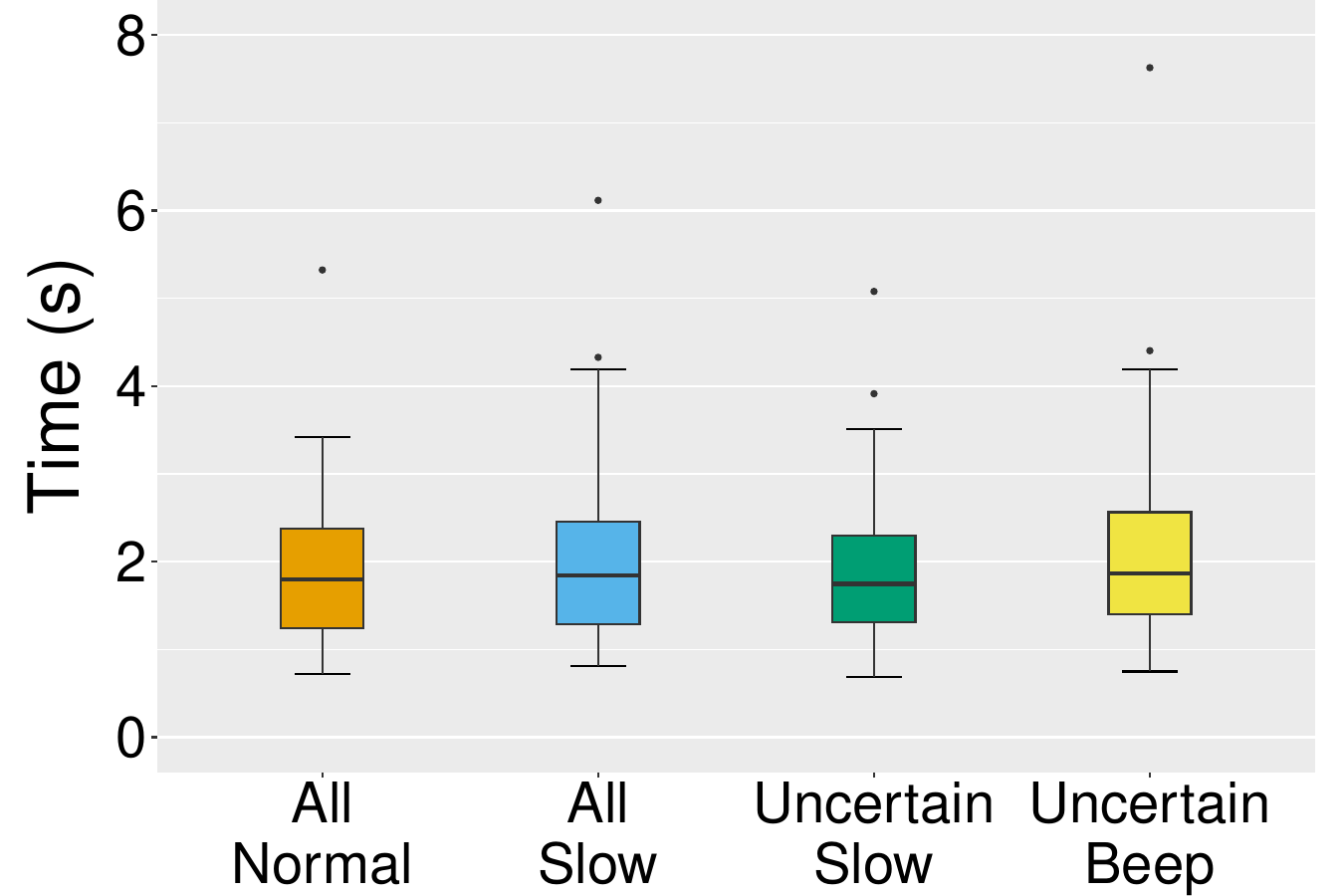} 
 \caption{Detection Time}
 \label{fig:time}
  \end{subfigure}%
    \caption{Comparison of the detection accuracy (top) and the detection time (bottom) in the four different study conditions.}
    \label{fig:boxplot}
\Description{Two box plots showing results in the four conditions. The first box plot compares the detection accuracy percentages. The second box plot compared the average detection times in seconds.}
\end{figure}

\subsection{Error Detection Accuracy}
We calculated how often users correctly determined whether the recognition result contained any errors (i.e.~by selecting yes or no after hearing the audio). As shown in Figure \ref{fig:acc}, the proportion of correct error detection was higher in \textsc{UncertainSlow} (85\%) compared to \textsc{AllNormal} (80\%), \textsc{AllSlow} (76\%), and \textsc{UncertainBeep} (79\%). A non-parametric ART test revealed a significant difference ($F_{3,141}=4.48$, $\eta_{p}^{2}=0.087$, $p= 0.005$). Post-hoc pairwise comparisons with Bonferroni correction found a significant difference between the \textsc{UncertainSlow} and \textsc{AllSlow} conditions ($p=0.002$). However, no significant differences were observed between \textsc{UncertainSlow} and \textsc{AllNormal} ($p=0.1$), \textsc{UncertainSlow} and \textsc{UncertainBeep} ($p=0.2$), \textsc{AllNormal} and \textsc{AllSlow} ($p=0.9$), \textsc{AllNormal} and \textsc{UncertainBeep} ($p=1.0$), or \textsc{AllSlow} and \textsc{UncertainBeep} ($p=1.$).

In contrast to the previous study by Hong and Findlater \cite{hong_identifying_2018} that reported improved error detection with a slow speech rate, our study did not find a significant difference in error detection between other pairs. However, our results suggest that slowing down the audio playback only when necessary might help users to better detect the presence of errors compared to uniformly slowing down the playback. As shown in Figure \ref{fig:acc}, the variance in per-participant error detection performance was smaller in the \textsc{UncertainSlow} ($SD=11.5$) condition compared to \textsc{AllNormal} ($SD=12.9$), \textsc{AllSlow} ($SD=15.8$), and \textsc{UncertainBeep} ($SD=16.6$). This may indicate the confidence score-based slowing helped some users avoid substantially lower accuracy compared to the average.

Unfortunately, we lacked sufficient data to reliably analyze the impact of different audio annotations on participants' ability to identify the specific locations of the errors. This was due to not every participant experiencing a sufficient number of recognition errors in each condition. However, we were did conduct some analysis by aggregating errors across all conditions. We found users correctly located 49\% of all errors. Broken down by type of recognition error, they located 2\% of insertion errors, 49\% of substitution errors, and 62\% of deletion errors. Actual substitution and deletion errors were identified by aligning the reference and recognition transcripts using the Levenshtein distance algorithm \cite{levenshtein1966binary}. We determined actual insertion errors by manual review. 

Across all conditions, the ratio of locating errors was 48\% for challenging phrases and 52\% for common phrases. This indicates participants missed nearly half of the errors in the transcribed text and this was not strongly influenced by the difficultly of the sentence. In particular, participants struggled to identify insertions, suggesting that detecting and correcting added words may necessitate greater attention.

In our study, participants were presented with both common and challenging phrases. The WER was significantly higher at 17\% for challenging phrases compared to 12\% for common phrases ($r$ = -0.97, $p$ < 0.001). This suggests that our approach of using challenging phrases to elicit more recognition errors was effective. 

\begin{figure*}[tb]
    \includegraphics[width=13cm]{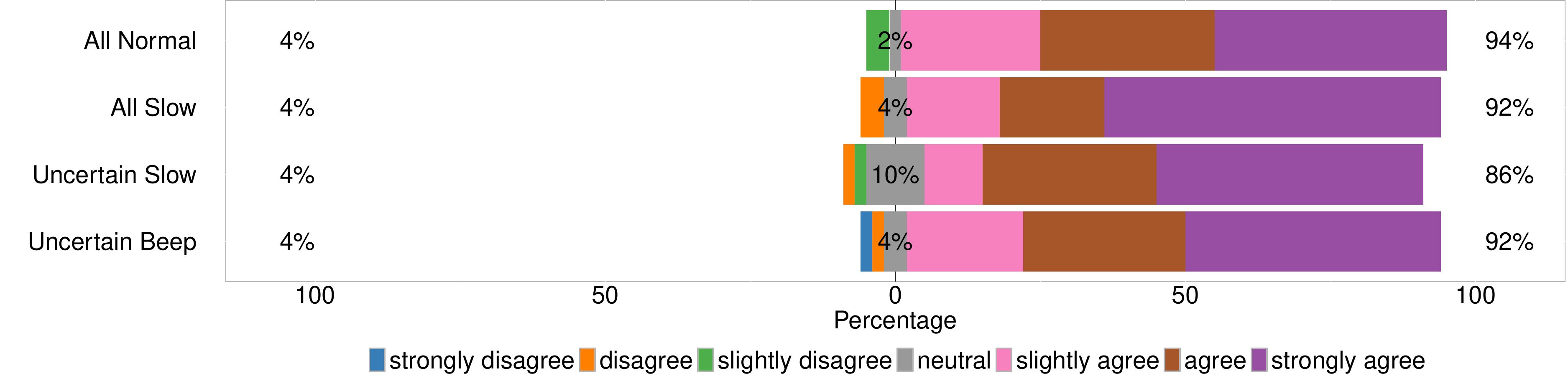}
    \caption{Participant feedback on how easy it was to identify errors in each condition. The percentages on the left are the portion of participants who strongly disagreed, disagreed, or slightly disagreed with the statements. The percentages in the middle correspond to the portion who were neutral. The percentages on the right correspond to those who strongly agreed, agreed, or slightly agreed.}
    \label{fig:feedback}
    \Description{A horizontal stacked bar plots showing participants' feedback on how easy it was to identify errors under different conditions. The percentages on the left represent the portion of participants who strongly disagreed or disagreed or slightly disagreed, the percentages in the middle correspond to neutral and the percentages on the right correspond to those who strongly agreed or agreed or slightly agreed with the statement.}
\end{figure*}

For common phrases, the proportion of correct error detection was higher in \textsc{UncertainSlow} (90.9\%) compared to \textsc{AllNormal} (84.7\%), \textsc{AllSlow} (83.5\%), and \textsc{UncertainBeep} (82.5\%). A non-parametric ART test revealed a significant difference ($F_{3,141}=9.40$, $\eta_p^2=0.167$, $p<0.001$). Post-hoc pairwise comparisons with Bonferroni correction found significant differences between \textsc{AllSlow} and \textsc{UncertainSlow} ($p=0.0003$), and between \textsc{AllSlow} and \textsc{AllNormal} ($p=0.0001$). No significant differences were observed between \textsc{UncertainSlow} and \textsc{UncertainBeep} ($p=1.0$) or between \textsc{AllNormal} and \textsc{UncertainBeep} ($p=1.0$). 

For challenging phrases, the proportion of correct error detection was also higher in \textsc{UncertainSlow} (79.9\%) compared to \textsc{UncertainBeep} (75.9\%), \textsc{AllNormal} (75.6\%), and \textsc{AllSlow} (68.5\%). An ART test revealed a significant difference ($F_{3,141} = 11.52$, $\eta_p^2=0.197$, $p<0.001$). Post-hoc pairwise comparisons with Bonferroni correction found significant differences between \textsc{AllNormal} and \textsc{AllSlow} ($p=0.0002$), and between \textsc{AllNormal} and \textsc{UncertainBeep} ($p=0.0003$). No significant differences were observed between \textsc{UncertainSlow} and \textsc{UncertainBeep} ($p=0.2$), or between \textsc{UncertainSlow} and \textsc{AllNormal} ($p=0.07$). 

Comparing results between the common and challenging phrases, it is evident that error detection accuracy decreases for challenging phrases. However, the relative advantages of \textsc{UncertainSlow} remained consistent and had the highest detection accuracy across both common (90.9\%) and challenging (79.9\%) phrases. This suggests that selectively slowing playback as done in the \textsc{UncertainSlow} condition effectively supports error detection across varying task difficulties. 

\subsection{Detection Time}
We measured the \textit{detection time} from the end of the audio playback to the participant's response of yes or no. Average detection times were similar: 1.98 seconds in \textsc{AllNormal}, 2.03 seconds in \textsc{All Slow}, 1.86 seconds in \textsc{Uncertain Slow}, and 2.26 seconds in \textsc{Uncertain Beep} condition (Figure \ref{fig:time}). These differences were not significant ($F_{3,141}=0.69$, $\eta_{p}^{2}=0.014$, $p=0.56$). 

We also calculated  the \textit{total time} it took participants to listen to the audio and to respond yes or no. The total time was 4.52\,s in \textsc{AllNormal}, 4.85\,s in \textsc{UncertainSlow}, 5.43\,s in \textsc{AllSlow}, and 5.42\,s in \textsc{UncertainBeep}. As expected given the similar detection times we observed between conditions, playing at the normal rate with no beep was the fastest. Always slowing the playback or adding the beep for uncertain recognitions resulted in a 20\% increase in total time compared to normal playback. Using a slower speaking rate for uncertain results instead of the beep was more time efficient, resulting in only a 7\% increase in total time compared to normal playback and an 11\% decrease compared to always slowing down the audio.

\subsection{Subjective Feedback}
Participants rated how easy it was to identify errors under four conditions on a 7-point Likert scale (1\hspace{0.1mm}=\hspace{0.1mm}strongly disagree, 7\hspace{0.1mm}=\hspace{0.1mm}strongly agree). As shown in Figure \ref{fig:feedback}, 94\% of participants found \textsc{AllNormal} easy, followed closely by 92\% for \textsc{AllSlow} and \textsc{UncertainBeep}, and 86\% for \textsc{UncertainSlow}. The Friedman test indicated no significant difference in participants' ratings across the four conditions ($\chi^2(3)= 2.94$, $p=0.40$).

Participants also rated whether they could anticipate when a sentence was likely to result in a recognition error. A majority felt they could anticipate errors based just on the sentence with 62\% expressing some level of agreement (slightly agree, agree, or strongly agree). In contrast, 24\% of participants expressed some level of disagreement (slightly disagree, disagree, or strongly disagree). The remaining 14\% were neutral.

\section{Discussion}
In our study, we used four audio annotations to assess participants' ability to detect errors in their transcribed speech. Our result suggests that using the recognizer's confidence in its results to change how we present the result audio can help users detect errors. 

One limitation of our study is we only considered native English speakers who were sighted. Blind users, for example, may have more experience listening to TTS, which could impact their ability to detect errors in TTS audio. Additionally, non-native speakers with ascents may have different experiences with speech recognition technology. Future research should explore how diverse populations detect errors when using speech recognition.

We do not know the environment our crowdsourced participants completed our study, but it is likely many were in a quiet environment. In real-world use, speech recognition users may be exposed to various types of noise and distractions that could affect their ability to detect errors. Moreover, users may be engaged in other tasks while using voice assistants (e.g.~ driving or exercising), which could also affect their ability to detect errors. Future studies could investigate how different contexts impact users' ability to detect speech recognition errors.

Our evaluation used a single static confidence threshold determined by our pilot testing. Instead, a system could try and dynamically adjust a user's threshold based on observing their interactions with previous recognition results. For example, if a result was above the threshold but the user corrected it, this may signal that a lower threshold is needed.

To create a realistic task, we had participants record themselves speaking sentences and used a state-of-the-art speech recognizer to present their actual transcription results. While we could have artificially forced recognition errors, we wanted to study the impact of our interventions in a more externally valid task with accurate speech recognition, but with some phrases containing difficult words (as might happen in real-world use). However, this resulted in our data lacking sufficient recognition errors for each participant and condition for fine-grained analysis of users' ability to locate errors. We suggest future work consider additional ways to ensure sufficient errors in each condition such as: 1) a longer or multi-session study with more utterances per condition, 2) adding noise to participant's audio to increase errors, or 3) occasionally presenting the second best recognition result.

Our experimental application first recorded a participant's entire spoken utterance before sending it to a remote server for speech recognition. Once the client received the recognition result, it was sent to another remote server to generate the TTS audio. This resulted in participants waiting around four seconds to hear their recognition result. A real-world interface could reduce this latency by: 1) streaming audio to the speech recognizer, 2) performing speech recognition and TTS on the same server, and 3) streaming TTS audio to the client as it is generated. 

We used the same TTS voice for all conditions. Future work could explore the impact of the specific TTS voice or technology (e.g.~concatenative versus neural TTS). Since synthetic speech can be generated in a hyperarticulate style \cite{aylett2005synthesising}, it would be interesting to investigate using hyperarticulate speech for low-confidence results. This might both help draw attention to the potential error and help users better discriminate true errors from false positives. 

We changed the presentation of the entire recognition result based on the utterance confidence score. If word-level confidence scores are available, future work could test modifying the audio of individual words suspected of being incorrect. This could be done by changing a word's speaking style, rate, or by adding auditory markers near the word. Another possibility would be to repeat suspected word errors to help users verify if they are true errors.

\section{Conclusion}
Our study investigated whether changing the audio of a speech recognition result based on its confidence score helped participants detect more recognition errors. We changed the audio either by slowly the TTS or by adding a beep tone. We found using confidence scores showed a trend toward improved error detection including reducing the variability in the detection accuracy of our users. However, the 85\% detection accuracy of selectively slowing playback was not statistically different than the 80\% accuracy of the baseline condition that played all results at normal speed. This highlights the need for further research to validate and expand upon these findings. Nevertheless, we believe our results can help inform the design of more effective error detection features for eyes-free speech interaction.

\begin{acks}
This material is based upon work supported by the NSF under Grant No. IIS-1909248.
\end{acks}

\bibliographystyle{ACM-Reference-Format}
\bibliography{ref}


\begin{thebibliography}{32}


\ifx \showCODEN    \undefined \def \showCODEN     #1{\unskip}     \fi
\ifx \showDOI      \undefined \def \showDOI       #1{#1}\fi
\ifx \showISBNx    \undefined \def \showISBNx     #1{\unskip}     \fi
\ifx \showISBNxiii \undefined \def \showISBNxiii  #1{\unskip}     \fi
\ifx \showISSN     \undefined \def \showISSN      #1{\unskip}     \fi
\ifx \showLCCN     \undefined \def \showLCCN      #1{\unskip}     \fi
\ifx \shownote     \undefined \def \shownote      #1{#1}          \fi
\ifx \showarticletitle \undefined \def \showarticletitle #1{#1}   \fi
\ifx \showURL      \undefined \def \showURL       {\relax}        \fi
\providecommand\bibfield[2]{#2}
\providecommand\bibinfo[2]{#2}
\providecommand\natexlab[1]{#1}
\providecommand\showeprint[2][]{arXiv:#2}

\bibitem[Aylett(2005)]%
        {aylett2005synthesising}
\bibfield{author}{\bibinfo{person}{Matthew~P Aylett}.} \bibinfo{year}{2005}\natexlab{}.
\newblock \showarticletitle{Synthesising hyperarticulation in unit selection TTS}. In \bibinfo{booktitle}{\emph{INTERSPEECH 2005-Eurospeech, 9th European Conference on Speech Communication and Technology, Lisbon, Portugal, September 4-8, 2005}}. \bibinfo{pages}{2521--2524}.
\newblock


\bibitem[Azenkot and Lee(2013)]%
        {azenkot_exploring_2013}
\bibfield{author}{\bibinfo{person}{Shiri Azenkot} {and} \bibinfo{person}{Nicole~B. Lee}.} \bibinfo{year}{2013}\natexlab{}.
\newblock \showarticletitle{Exploring the use of speech input by blind people on mobile devices}. In \bibinfo{booktitle}{\emph{Proceedings of the 15th {International} {ACM} {SIGACCESS} {Conference} on {Computers} and {Accessibility}}}. \bibinfo{publisher}{ACM}, \bibinfo{address}{Bellevue Washington}, \bibinfo{pages}{1--8}.
\newblock
\showISBNx{978-1-4503-2405-2}
\urldef\tempurl%
\url{https://doi.org/10.1145/2513383.2513440}
\showDOI{\tempurl}


\bibitem[Baevski et~al\mbox{.}(2020)]%
        {baevski_wav2vec_2020}
\bibfield{author}{\bibinfo{person}{Alexei Baevski}, \bibinfo{person}{Henry Zhou}, \bibinfo{person}{Abdelrahman Mohamed}, {and} \bibinfo{person}{Michael Auli}.} \bibinfo{year}{2020}\natexlab{}.
\newblock \showarticletitle{wav2vec 2.0: a framework for self-supervised learning of speech representations}. In \bibinfo{booktitle}{\emph{Proceedings of the 34th {International} {Conference} on {Neural} {Information} {Processing} {Systems}}} \emph{(\bibinfo{series}{{NIPS}'20})}. \bibinfo{publisher}{Curran Associates Inc.}, \bibinfo{address}{Red Hook, NY, USA}, \bibinfo{pages}{12449--12460}.
\newblock
\showISBNx{978-1-71382-954-6}


\bibitem[Berke et~al\mbox{.}(2017)]%
        {berke_captions}
\bibfield{author}{\bibinfo{person}{Larwan Berke}, \bibinfo{person}{Christopher Caulfield}, {and} \bibinfo{person}{Matt Huenerfauth}.} \bibinfo{year}{2017}\natexlab{}.
\newblock \showarticletitle{Deaf and Hard-of-Hearing Perspectives on Imperfect Automatic Speech Recognition for Captioning One-on-One Meetings} \emph{(\bibinfo{series}{ASSETS '17})}. \bibinfo{publisher}{Association for Computing Machinery}, \bibinfo{address}{New York, NY, USA}, \bibinfo{pages}{155–164}.
\newblock
\showISBNx{9781450349260}
\urldef\tempurl%
\url{https://doi.org/10.1145/3132525.3132541}
\showDOI{\tempurl}


\bibitem[Bilmes et~al\mbox{.}(2005)]%
        {bilmes_vocal_2005}
\bibfield{author}{\bibinfo{person}{Jeff~A. Bilmes}, \bibinfo{person}{Patricia Dowden}, \bibinfo{person}{Howard Chizeck}, \bibinfo{person}{Xiao Li}, \bibinfo{person}{Jonathan Malkin}, \bibinfo{person}{Kelley Kilanski}, \bibinfo{person}{Richard Wright}, \bibinfo{person}{Katrin Kirchhoff}, \bibinfo{person}{Amarnag Subramanya}, \bibinfo{person}{Susumu Harada}, {and} \bibinfo{person}{James~A. Landay}.} \bibinfo{year}{2005}\natexlab{}.
\newblock \showarticletitle{The vocal joystick: a voice-based human-computer interface for individuals with motor impairments}. In \bibinfo{booktitle}{\emph{Proceedings of the conference on {Human} {Language} {Technology} and {Empirical} {Methods} in {Natural} {Language} {Processing} - {HLT} '05}}. \bibinfo{publisher}{Association for Computational Linguistics}, \bibinfo{address}{Vancouver, British Columbia, Canada}, \bibinfo{pages}{995--1002}.
\newblock
\urldef\tempurl%
\url{https://doi.org/10.3115/1220575.1220700}
\showDOI{\tempurl}


\bibitem[Burke et~al\mbox{.}(2006)]%
        {burke_correction_2006}
\bibfield{author}{\bibinfo{person}{Moira Burke}, \bibinfo{person}{Brian Amento}, {and} \bibinfo{person}{Philip Isenhour}.} \bibinfo{year}{2006}\natexlab{}.
\newblock \showarticletitle{Error Correction of Voicemail Transcripts in SCANMail}. In \bibinfo{booktitle}{\emph{Proceedings of the SIGCHI Conference on Human Factors in Computing Systems}} (Montr\'{e}al, Qu\'{e}bec, Canada) \emph{(\bibinfo{series}{CHI '06})}. \bibinfo{publisher}{Association for Computing Machinery}, \bibinfo{address}{New York, NY, USA}, \bibinfo{pages}{339–348}.
\newblock
\showISBNx{1595933727}
\urldef\tempurl%
\url{https://doi.org/10.1145/1124772.1124823}
\showDOI{\tempurl}


\bibitem[Cohn and Zellou(2020)]%
        {cohn_perception}
\bibfield{author}{\bibinfo{person}{Michelle Cohn} {and} \bibinfo{person}{Georgia Zellou}.} \bibinfo{year}{2020}\natexlab{}.
\newblock \showarticletitle{Perception of concatenative vs. neural text-to-speech (TTS): Differences in intelligibility in noise and language attitudes}. In \bibinfo{booktitle}{\emph{Proceedings of Interspeech}}.
\newblock


\bibitem[Fan et~al\mbox{.}(2021)]%
        {fan_just_2021}
\bibfield{author}{\bibinfo{person}{Jiayue Fan}, \bibinfo{person}{Chenning Xu}, \bibinfo{person}{Chun Yu}, {and} \bibinfo{person}{Yuanchun Shi}.} \bibinfo{year}{2021}\natexlab{}.
\newblock \showarticletitle{Just {Speak} {It}: {Minimize} {Cognitive} {Load} for {Eyes}-{Free} {Text} {Editing} with a {Smart} {Voice} {Assistant}}. In \bibinfo{booktitle}{\emph{The 34th {Annual} {ACM} {Symposium} on {User} {Interface} {Software} and {Technology}}}. \bibinfo{publisher}{ACM}, \bibinfo{address}{Virtual Event USA}, \bibinfo{pages}{910--921}.
\newblock
\showISBNx{978-1-4503-8635-7}
\urldef\tempurl%
\url{https://doi.org/10.1145/3472749.3474795}
\showDOI{\tempurl}


\bibitem[Fujiwara(2016)]%
        {fuji_correction_2016}
\bibfield{author}{\bibinfo{person}{Kazuki Fujiwara}.} \bibinfo{year}{2016}\natexlab{}.
\newblock \showarticletitle{Error Correction of Speech Recognition by Custom Phonetic Alphabet Input for Ultra-Small Devices}. In \bibinfo{booktitle}{\emph{Proceedings of the 2016 CHI Conference Extended Abstracts on Human Factors in Computing Systems}} (San Jose, California, USA) \emph{(\bibinfo{series}{CHI EA '16})}. \bibinfo{publisher}{Association for Computing Machinery}, \bibinfo{address}{New York, NY, USA}, \bibinfo{pages}{104–109}.
\newblock
\showISBNx{9781450340823}
\urldef\tempurl%
\url{https://doi.org/10.1145/2851581.2890380}
\showDOI{\tempurl}


\bibitem[Ghosh et~al\mbox{.}(2020)]%
        {ghosh_commanding_2020}
\bibfield{author}{\bibinfo{person}{Debjyoti Ghosh}, \bibinfo{person}{Can Liu}, \bibinfo{person}{Shengdong Zhao}, {and} \bibinfo{person}{Kotaro Hara}.} \bibinfo{year}{2020}\natexlab{}.
\newblock \showarticletitle{Commanding and {Re}-{Dictation}: {Developing} {Eyes}-{Free} {Voice}-{Based} {Interaction} for {Editing} {Dictated} {Text}}.
\newblock \bibinfo{journal}{\emph{ACM Transactions on Computer-Human Interaction}} \bibinfo{volume}{27}, \bibinfo{number}{4} (\bibinfo{date}{Aug.} \bibinfo{year}{2020}), \bibinfo{pages}{1--31}.
\newblock
\showISSN{1073-0516, 1557-7325}
\urldef\tempurl%
\url{https://doi.org/10.1145/3390889}
\showDOI{\tempurl}


\bibitem[Gillick et~al\mbox{.}(1997)]%
        {gillick_probabilistic_1997}
\bibfield{author}{\bibinfo{person}{L. Gillick}, \bibinfo{person}{Y. Ito}, {and} \bibinfo{person}{J. Young}.} \bibinfo{year}{1997}\natexlab{}.
\newblock \showarticletitle{A probabilistic approach to confidence estimation and evaluation}. In \bibinfo{booktitle}{\emph{1997 {IEEE} {International} {Conference} on {Acoustics}, {Speech}, and {Signal} {Processing}}}, Vol.~\bibinfo{volume}{2}. \bibinfo{pages}{879--882 vol.2}.
\newblock
\urldef\tempurl%
\url{https://doi.org/10.1109/ICASSP.1997.596076}
\showDOI{\tempurl}
\newblock
\shownote{ISSN: 1520-6149}.


\bibitem[Goldwater et~al\mbox{.}(2010)]%
        {goldwater_which_2010}
\bibfield{author}{\bibinfo{person}{Sharon Goldwater}, \bibinfo{person}{Dan Jurafsky}, {and} \bibinfo{person}{Christopher~D. Manning}.} \bibinfo{year}{2010}\natexlab{}.
\newblock \showarticletitle{Which words are hard to recognize? {Prosodic}, lexical, and disfluency factors that increase speech recognition error rates}.
\newblock \bibinfo{journal}{\emph{Speech Communication}} \bibinfo{volume}{52}, \bibinfo{number}{3} (\bibinfo{date}{March} \bibinfo{year}{2010}), \bibinfo{pages}{181--200}.
\newblock
\showISSN{0167-6393}
\urldef\tempurl%
\url{https://doi.org/10.1016/j.specom.2009.10.001}
\showDOI{\tempurl}


\bibitem[Hong and Findlater(2018)]%
        {hong_identifying_2018}
\bibfield{author}{\bibinfo{person}{Jonggi Hong} {and} \bibinfo{person}{Leah Findlater}.} \bibinfo{year}{2018}\natexlab{}.
\newblock \showarticletitle{Identifying {Speech} {Input} {Errors} {Through} {Audio}-{Only} {Interaction}}. In \bibinfo{booktitle}{\emph{Proceedings of the 2018 {CHI} {Conference} on {Human} {Factors} in {Computing} {Systems}}} \emph{(\bibinfo{series}{{CHI} '18})}. \bibinfo{publisher}{Association for Computing Machinery}, \bibinfo{address}{New York, NY, USA}, \bibinfo{pages}{1--12}.
\newblock
\showISBNx{978-1-4503-5620-6}
\urldef\tempurl%
\url{https://doi.org/10.1145/3173574.3174141}
\showDOI{\tempurl}


\bibitem[Hong et~al\mbox{.}(2020)]%
        {hong_review_2020}
\bibfield{author}{\bibinfo{person}{Jonggi Hong}, \bibinfo{person}{Christine Vaing}, \bibinfo{person}{Hernisa Kacorri}, {and} \bibinfo{person}{Leah Findlater}.} \bibinfo{year}{2020}\natexlab{}.
\newblock \showarticletitle{Reviewing Speech Input with Audio: Differences between Blind and Sighted Users}.
\newblock \bibinfo{journal}{\emph{ACM Trans. Access. Comput.}} \bibinfo{volume}{13}, \bibinfo{number}{1}, Article \bibinfo{articleno}{2} (\bibinfo{date}{apr} \bibinfo{year}{2020}), \bibinfo{numpages}{28}~pages.
\newblock
\showISSN{1936-7228}
\urldef\tempurl%
\url{https://doi.org/10.1145/3382039}
\showDOI{\tempurl}


\bibitem[Huang et~al\mbox{.}(2022)]%
        {huang_duiva_2022}
\bibfield{author}{\bibinfo{person}{Jizhou Huang}, \bibinfo{person}{Haifeng Wang}, \bibinfo{person}{Shiqiang Ding}, {and} \bibinfo{person}{Shaolei Wang}.} \bibinfo{year}{2022}\natexlab{}.
\newblock \showarticletitle{{DuIVA}: {An} {Intelligent} {Voice} {Assistant} for {Hands}-free and {Eyes}-free {Voice} {Interaction} with the {Baidu} {Maps} {App}}. In \bibinfo{booktitle}{\emph{Proceedings of the 28th {ACM} {SIGKDD} {Conference} on {Knowledge} {Discovery} and {Data} {Mining}}}. \bibinfo{publisher}{ACM}, \bibinfo{address}{Washington DC USA}, \bibinfo{pages}{3040--3050}.
\newblock
\showISBNx{978-1-4503-9385-0}
\urldef\tempurl%
\url{https://doi.org/10.1145/3534678.3539030}
\showDOI{\tempurl}


\bibitem[Levenshtein et~al\mbox{.}(1966)]%
        {levenshtein1966binary}
\bibfield{author}{\bibinfo{person}{Vladimir~I Levenshtein} {et~al\mbox{.}}} \bibinfo{year}{1966}\natexlab{}.
\newblock \showarticletitle{Binary codes capable of correcting deletions, insertions, and reversals}. In \bibinfo{booktitle}{\emph{Soviet physics doklady}}, Vol.~\bibinfo{volume}{10}. Soviet Union, \bibinfo{pages}{707--710}.
\newblock


\bibitem[Liu et~al\mbox{.}(2023)]%
        {liu_unsupervised_2022}
\bibfield{author}{\bibinfo{person}{Alexander~H. Liu}, \bibinfo{person}{Wei-Ning Hsu}, \bibinfo{person}{Michael Auli}, {and} \bibinfo{person}{Alexei Baevski}.} \bibinfo{year}{2023}\natexlab{}.
\newblock \showarticletitle{Towards End-to-End Unsupervised Speech Recognition}. In \bibinfo{booktitle}{\emph{2022 IEEE Spoken Language Technology Workshop (SLT)}}. \bibinfo{pages}{221--228}.
\newblock
\urldef\tempurl%
\url{https://doi.org/10.1109/SLT54892.2023.10023187}
\showDOI{\tempurl}


\bibitem[Metatla et~al\mbox{.}(2019)]%
        {metatla_voice_2019}
\bibfield{author}{\bibinfo{person}{Oussama Metatla}, \bibinfo{person}{Alison Oldfield}, \bibinfo{person}{Taimur Ahmed}, \bibinfo{person}{Antonis Vafeas}, {and} \bibinfo{person}{Sunny Miglani}.} \bibinfo{year}{2019}\natexlab{}.
\newblock \showarticletitle{Voice {User} {Interfaces} in {Schools}: {Co}-designing for {Inclusion} with {Visually}-{Impaired} and {Sighted} {Pupils}}. In \bibinfo{booktitle}{\emph{Proceedings of the 2019 {CHI} {Conference} on {Human} {Factors} in {Computing} {Systems}}}. \bibinfo{publisher}{ACM}, \bibinfo{address}{Glasgow Scotland Uk}, \bibinfo{pages}{1--15}.
\newblock
\showISBNx{978-1-4503-5970-2}
\urldef\tempurl%
\url{https://doi.org/10.1145/3290605.3300608}
\showDOI{\tempurl}


\bibitem[Nowrin et~al\mbox{.}(2022)]%
        {nowrin_exploring_2022}
\bibfield{author}{\bibinfo{person}{Sadia Nowrin}, \bibinfo{person}{Patricia OrdóñEz}, {and} \bibinfo{person}{Keith Vertanen}.} \bibinfo{year}{2022}\natexlab{}.
\newblock \showarticletitle{Exploring {Motor}-impaired {Programmers}’ {Use} of {Speech} {Recognition}}. In \bibinfo{booktitle}{\emph{The 24th {International} {ACM} {SIGACCESS} {Conference} on {Computers} and {Accessibility}}}. \bibinfo{publisher}{ACM}, \bibinfo{address}{Athens Greece}, \bibinfo{pages}{1--4}.
\newblock
\showISBNx{978-1-4503-9258-7}
\urldef\tempurl%
\url{https://doi.org/10.1145/3517428.3550392}
\showDOI{\tempurl}


\bibitem[Noyes and Frankish(1994)]%
        {noyes_errors}
\bibfield{author}{\bibinfo{person}{JM Noyes} {and} \bibinfo{person}{CR Frankish}.} \bibinfo{year}{1994}\natexlab{}.
\newblock \showarticletitle{Errors and error correction in automatic speech recognition systems}.
\newblock \bibinfo{journal}{\emph{Ergonomics}} \bibinfo{volume}{37}, \bibinfo{number}{11} (\bibinfo{year}{1994}), \bibinfo{pages}{1943--1957}.
\newblock


\bibitem[Piquard-Kipffer et~al\mbox{.}(2015)]%
        {piquard_qualitative}
\bibfield{author}{\bibinfo{person}{Agn{\`e}s Piquard-Kipffer}, \bibinfo{person}{Odile Mella}, \bibinfo{person}{J{\'e}r{\'e}my Miranda}, \bibinfo{person}{Denis Jouvet}, {and} \bibinfo{person}{Luiza Orosanu}.} \bibinfo{year}{2015}\natexlab{}.
\newblock \showarticletitle{Qualitative investigation of the display of speech recognition results for communication with deaf people}. In \bibinfo{booktitle}{\emph{Proceedings of SLPAT 2015: 6th Workshop on Speech and Language Processing for Assistive Technologies}}. \bibinfo{pages}{36--41}.
\newblock


\bibitem[Pradhan et~al\mbox{.}(2018)]%
        {pradhan_accessibility_2018}
\bibfield{author}{\bibinfo{person}{Alisha Pradhan}, \bibinfo{person}{Kanika Mehta}, {and} \bibinfo{person}{Leah Findlater}.} \bibinfo{year}{2018}\natexlab{}.
\newblock \showarticletitle{"{Accessibility} {Came} by {Accident}": {Use} of {Voice}-{Controlled} {Intelligent} {Personal} {Assistants} by {People} with {Disabilities}}. In \bibinfo{booktitle}{\emph{Proceedings of the 2018 {CHI} {Conference} on {Human} {Factors} in {Computing} {Systems}}} \emph{(\bibinfo{series}{{CHI} '18})}. \bibinfo{publisher}{Association for Computing Machinery}, \bibinfo{address}{New York, NY, USA}, \bibinfo{pages}{1--13}.
\newblock
\showISBNx{978-1-4503-5620-6}
\urldef\tempurl%
\url{https://doi.org/10.1145/3173574.3174033}
\showDOI{\tempurl}


\bibitem[Price and Sears(2005)]%
        {price_correct_2005}
\bibfield{author}{\bibinfo{person}{{Kathleen J.} Price} {and} \bibinfo{person}{Andrew Sears}.} \bibinfo{year}{2005}\natexlab{}.
\newblock \showarticletitle{Speech-based text entry for mobile handheld devices: An analysis of efficacy and error correction techniques for server-based solutions}.
\newblock \bibinfo{journal}{\emph{International Journal of Human-Computer Interaction}} \bibinfo{volume}{19}, \bibinfo{number}{3} (\bibinfo{year}{2005}), \bibinfo{pages}{279--304}.
\newblock
\showISSN{1044-7318}
\urldef\tempurl%
\url{https://doi.org/10.1207/s15327590ijhc1903_1}
\showDOI{\tempurl}
\newblock
\shownote{Funding Information: The authors thank Aether Systems, Inc. for their support of this research. This material is based upon work supported by the National Science Foundation under Grants IIS–9910607 and IIS–0121570. Any opinions, findings and conclusions or recommendations expressed in this material are those of the authors and do not necessarily reflect the views of the National Science Foundation (NSF).}.


\bibitem[Shiver and Wolfe(2015)]%
        {shiver_evaluating}
\bibfield{author}{\bibinfo{person}{Brent~N Shiver} {and} \bibinfo{person}{Rosalee~J Wolfe}.} \bibinfo{year}{2015}\natexlab{}.
\newblock \showarticletitle{Evaluating alternatives for better deaf accessibility to selected web-based multimedia}. In \bibinfo{booktitle}{\emph{Proceedings of the 17th international ACM SIGACCESS conference on computers \& accessibility}}. \bibinfo{pages}{231--238}.
\newblock


\bibitem[Soe et~al\mbox{.}(2021)]%
        {htut_subtitling}
\bibfield{author}{\bibinfo{person}{Than~Htut Soe}, \bibinfo{person}{Frode Guribye}, {and} \bibinfo{person}{Marija Slavkovik}.} \bibinfo{year}{2021}\natexlab{}.
\newblock \showarticletitle{Evaluating AI assisted subtitling}. In \bibinfo{booktitle}{\emph{Proceedings of the 2021 ACM International Conference on Interactive Media Experiences}} (Virtual Event, USA) \emph{(\bibinfo{series}{IMX '21})}. \bibinfo{publisher}{Association for Computing Machinery}, \bibinfo{address}{New York, NY, USA}, \bibinfo{pages}{96–107}.
\newblock
\showISBNx{9781450383899}
\urldef\tempurl%
\url{https://doi.org/10.1145/3452918.3458792}
\showDOI{\tempurl}


\bibitem[Vertanen et~al\mbox{.}(2019)]%
        {vertanen_velociwatch_2019}
\bibfield{author}{\bibinfo{person}{Keith Vertanen}, \bibinfo{person}{Dylan Gaines}, \bibinfo{person}{Crystal Fletcher}, \bibinfo{person}{Alex~M. Stanage}, \bibinfo{person}{Robbie Watling}, {and} \bibinfo{person}{Per~Ola Kristensson}.} \bibinfo{year}{2019}\natexlab{}.
\newblock \showarticletitle{{VelociWatch}: {Designing} and {Evaluating} a {Virtual} {Keyboard} for the {Input} of {Challenging} {Text}}. In \bibinfo{booktitle}{\emph{Proceedings of the 2019 {CHI} {Conference} on {Human} {Factors} in {Computing} {Systems}}}. \bibinfo{publisher}{ACM}, \bibinfo{address}{Glasgow Scotland Uk}, \bibinfo{pages}{1--14}.
\newblock
\showISBNx{978-1-4503-5970-2}
\urldef\tempurl%
\url{https://doi.org/10.1145/3290605.3300821}
\showDOI{\tempurl}


\bibitem[Vertanen and Kristensson(2008)]%
        {vertanen_chi2008}
\bibfield{author}{\bibinfo{person}{Keith Vertanen} {and} \bibinfo{person}{Per~Ola Kristensson}.} \bibinfo{year}{2008}\natexlab{}.
\newblock \showarticletitle{On the benefits of confidence visualization in speech recognition}. In \bibinfo{booktitle}{\emph{CHI '08: Proceedings of the SIGCHI conference on Human Factors in computing systems}} (Florence, Italy). \bibinfo{publisher}{ACM}, \bibinfo{pages}{1497--1500}.
\newblock


\bibitem[Vertanen and Kristensson(2009)]%
        {vertanen_automatic_2009}
\bibfield{author}{\bibinfo{person}{Keith Vertanen} {and} \bibinfo{person}{Per~Ola Kristensson}.} \bibinfo{year}{2009}\natexlab{}.
\newblock \showarticletitle{Automatic selection of recognition errors by respeaking the intended text}. In \bibinfo{booktitle}{\emph{2009 {IEEE} {Workshop} on {Automatic} {Speech} {Recognition} \& {Understanding}}}. \bibinfo{pages}{130--135}.
\newblock
\urldef\tempurl%
\url{https://doi.org/10.1109/ASRU.2009.5373347}
\showDOI{\tempurl}


\bibitem[Vtyurina et~al\mbox{.}(2019)]%
        {vtyurina_bridging_2019}
\bibfield{author}{\bibinfo{person}{Alexandra Vtyurina}, \bibinfo{person}{Adam Fourney}, \bibinfo{person}{Meredith~Ringel Morris}, \bibinfo{person}{Leah Findlater}, {and} \bibinfo{person}{Ryen~W. White}.} \bibinfo{year}{2019}\natexlab{}.
\newblock \showarticletitle{Bridging {Screen} {Readers} and {Voice} {Assistants} for {Enhanced} {Eyes}-{Free} {Web} {Search}}. In \bibinfo{booktitle}{\emph{The {World} {Wide} {Web} {Conference}}}. \bibinfo{publisher}{ACM}, \bibinfo{address}{San Francisco CA USA}, \bibinfo{pages}{3590--3594}.
\newblock
\showISBNx{978-1-4503-6674-8}
\urldef\tempurl%
\url{https://doi.org/10.1145/3308558.3314136}
\showDOI{\tempurl}


\bibitem[Wagner et~al\mbox{.}(2012)]%
        {wagner_programming_2012}
\bibfield{author}{\bibinfo{person}{Amber Wagner}, \bibinfo{person}{Ramaraju Rudraraju}, \bibinfo{person}{Srinivasa Datla}, \bibinfo{person}{Avishek Banerjee}, \bibinfo{person}{Mandar Sudame}, {and} \bibinfo{person}{Jeff Gray}.} \bibinfo{year}{2012}\natexlab{}.
\newblock \showarticletitle{Programming by voice: a hands-free approach for motorically challenged children}. In \bibinfo{booktitle}{\emph{{CHI} '12 {Extended} {Abstracts} on {Human} {Factors} in {Computing} {Systems}}}. \bibinfo{publisher}{ACM}, \bibinfo{address}{Austin Texas USA}, \bibinfo{pages}{2087--2092}.
\newblock
\showISBNx{978-1-4503-1016-1}
\urldef\tempurl%
\url{https://doi.org/10.1145/2212776.2223757}
\showDOI{\tempurl}


\bibitem[Wang et~al\mbox{.}(2022)]%
        {wang_wav2vec-switch_2022}
\bibfield{author}{\bibinfo{person}{Yiming Wang}, \bibinfo{person}{Jinyu Li}, \bibinfo{person}{Heming Wang}, \bibinfo{person}{Yao Qian}, \bibinfo{person}{Chengyi Wang}, {and} \bibinfo{person}{Yu Wu}.} \bibinfo{year}{2022}\natexlab{}.
\newblock \showarticletitle{Wav2vec-{Switch}: {Contrastive} {Learning} from {Original}-{Noisy} {Speech} {Pairs} for {Robust} {Speech} {Recognition}}. In \bibinfo{booktitle}{\emph{{ICASSP} 2022 - 2022 {IEEE} {International} {Conference} on {Acoustics}, {Speech} and {Signal} {Processing} ({ICASSP})}}. \bibinfo{pages}{7097--7101}.
\newblock
\urldef\tempurl%
\url{https://doi.org/10.1109/ICASSP43922.2022.9746929}
\showDOI{\tempurl}
\newblock
\shownote{ISSN: 2379-190X}.


\bibitem[Weninger et~al\mbox{.}(2015)]%
        {weninger_speech_2015}
\bibfield{author}{\bibinfo{person}{Felix Weninger}, \bibinfo{person}{Hakan Erdogan}, \bibinfo{person}{Shinji Watanabe}, \bibinfo{person}{Emmanuel Vincent}, \bibinfo{person}{Jonathan Le~Roux}, \bibinfo{person}{John~R. Hershey}, {and} \bibinfo{person}{Björn Schuller}.} \bibinfo{year}{2015}\natexlab{}.
\newblock \showarticletitle{Speech {Enhancement} with {LSTM} {Recurrent} {Neural} {Networks} and its {Application} to {Noise}-{Robust} {ASR}}. In \bibinfo{booktitle}{\emph{Latent {Variable} {Analysis} and {Signal} {Separation}}} \emph{(\bibinfo{series}{Lecture {Notes} in {Computer} {Science}})}, \bibfield{editor}{\bibinfo{person}{Emmanuel Vincent}, \bibinfo{person}{Arie Yeredor}, \bibinfo{person}{Zbyněk Koldovský}, {and} \bibinfo{person}{Petr Tichavský}} (Eds.). \bibinfo{publisher}{Springer International Publishing}, \bibinfo{address}{Cham}, \bibinfo{pages}{91--99}.
\newblock
\showISBNx{978-3-319-22482-4}
\urldef\tempurl%
\url{https://doi.org/10.1007/978-3-319-22482-4_11}
\showDOI{\tempurl}


\end{thebibliography}

\appendix

\section{Questionnaire\label{sec_questionnaire}}
\begin{figure*}[tb]
\centering
\fbox{\includegraphics[width=14cm]{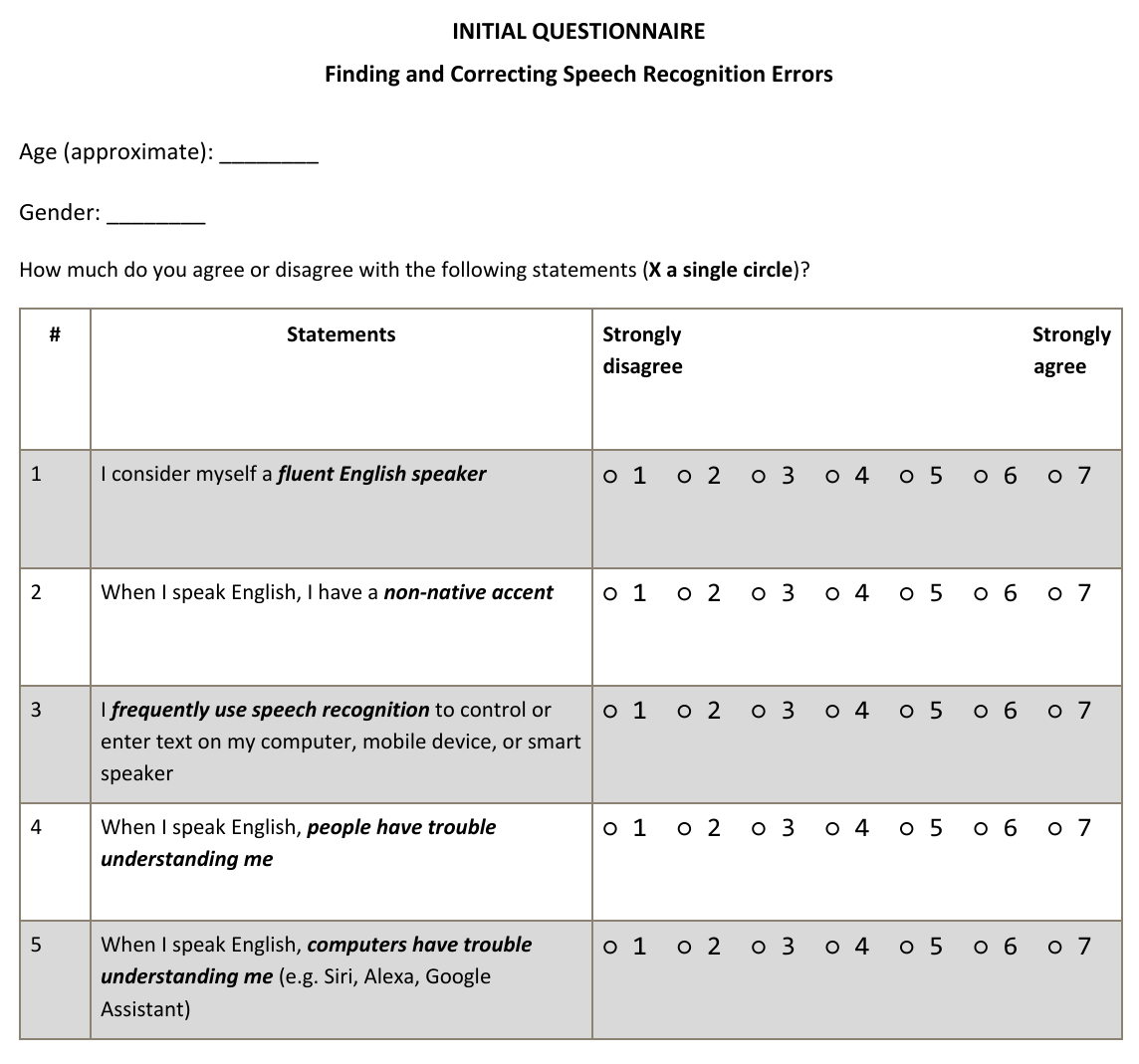}}
\caption{Our initial questionnaire asked participants their age and gender. They then rated five statements about their English ability and their experience with speech recognition. Statements were rated on a 7-point Likert scale.}
\label{questionnaire_init}
\Description{The initial questionnaire in which participants gave free form responses for their age and gender. They then rated the statements: 1) I consider myself a fluent English speaker, 2) When I speak English, I have have a non-native accent, 3) I frequently use speech recognition to control or enter text on my computer, mobile device, or smart speaker, 4) When I speak English, people have trouble understanding me, and 5) When I speak English, computers have trouble understanding me (e.g. Siri, Alexa, Google Assistant).}
\end{figure*}

\begin{figure*}[tb]
\centering
\fbox{\includegraphics[width=14cm]{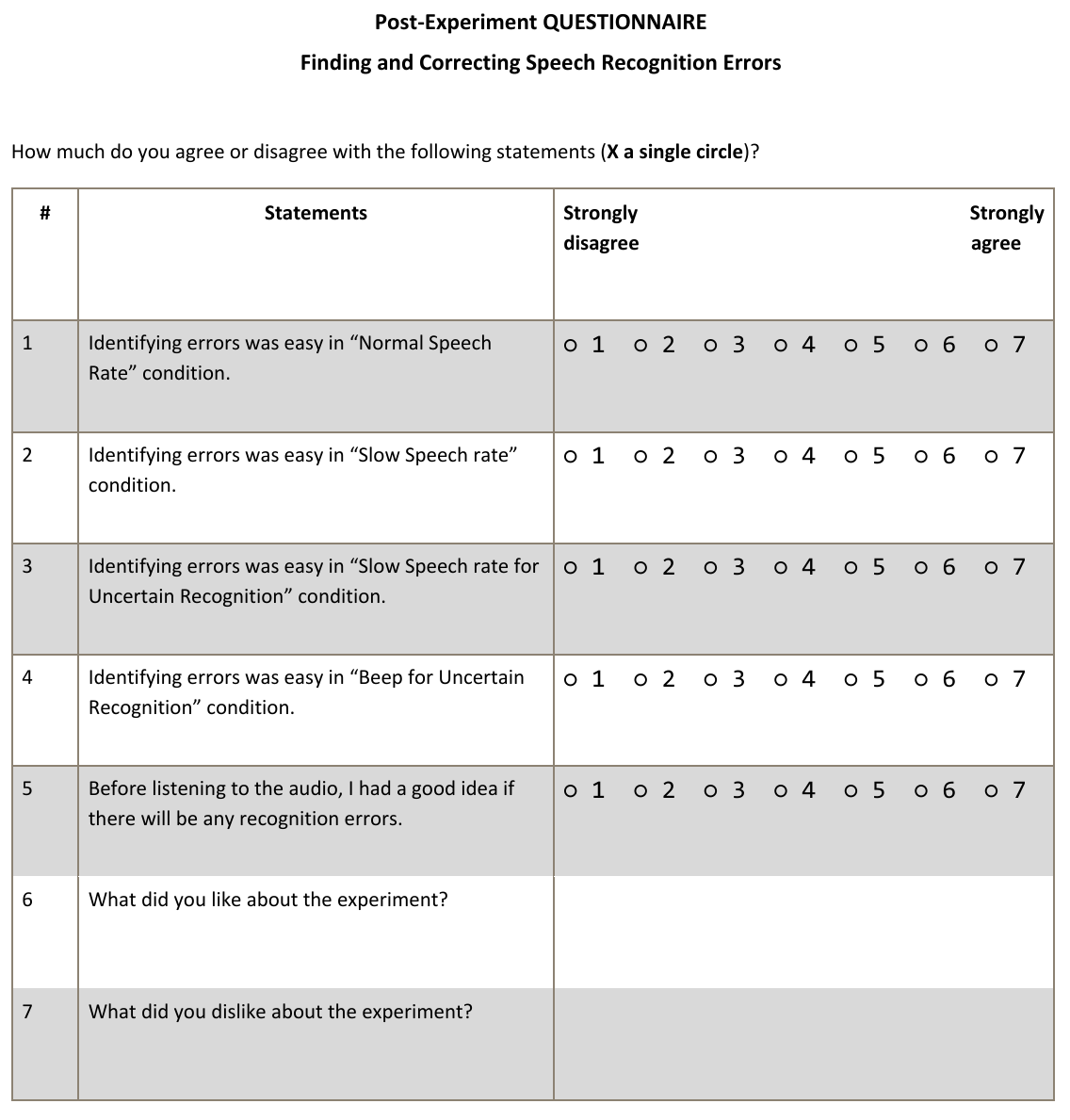}}
\caption{Our final questionnaire asked participants to rate five statements about their ability to find errors in the experiment's four conditions. They also rated their ability to anticipate sentences that would likely have recognition errors. Statements were rated on a 7-point Likert scale. We also asked open two open ended questions about what they liked or disliked about the experiment.}
\label{questionnaire_final}
\Description{The final questionnaire in which participants rated the statements: 1) Identifying errors was easy in Normal Speech Rate condition, 2) Identifying errors was easy in Slow Speech rate condition, 3) Identifying errors was easy in Slow Speech rate condition, 4) Identifying errors was easy in Beep for Uncertain Recognition condition, and 5) Before listening to the audio, I had a good idea if there will be any recognition errors. Finally, they answered two free form questions: 1) What did you like about the experiment, and 2) What did you dislike about the experiment.}
\end{figure*}

Figure \ref{questionnaire_init} shows the questions we asked participants at the start of the study. Figure \ref{questionnaire_final} show the questions we asked participants at the end of the study.

\end{document}